\documentclass[twocolumn,floats,prb,aps,showpacs]{revtex4}
\usepackage[final]{graphicx}
\usepackage[dvipsnames]{color}
\usepackage{dcolumn}
\usepackage{hhline}
\DeclareGraphicsExtensions{.pdf,.eps}

\begin{document}

\title{First-principles GW calculations for DNA and RNA nucleobases}

\author{Carina Faber$^{1,2}$, Claudio Attaccalite$^{1}$, Valerio Olevano$^{1}$, 
Erich Runge$^{2}$, Xavier Blase$^{1}$}

\affiliation{$^1$Institut N\'{e}el, CNRS and Universit\'{e} Joseph Fourier,
B.P. 166, 38042 Grenoble Cedex 09, France. \\ 
$^2$Institut f\"{u}r Physik, Technische Universit\"{a}t Ilmenau, 98693 Ilmenau, Germany. }

\date{\today}

\begin{abstract}
On the basis of first-principles GW calculations, we study the quasiparticle 
properties of the guanine, adenine, cytosine, thymine, and uracil DNA and RNA
nucleobases.  Beyond standard G$_0$W$_0$ calculations, starting from Kohn-Sham 
eigenstates obtained with (semi)local functionals, a simple self-consistency 
on the eigenvalues allows to obtain vertical ionization energies and electron affinities within
an average 0.11 eV and 0.18 eV error respectively as compared to state-of-the-art
coupled-cluster and multi-configurational perturbative quantum chemistry approaches.  
Further, GW calculations predict the correct $\pi$-character of the highest occupied state,
thanks to several level crossings between density functional and GW calculations.
Our study is based on a recent gaussian-basis implementation of GW with
explicit treatment of dynamical screening through contour deformation techniques.
\end{abstract}

\maketitle

\begin{figure*}
\begin{center}
\includegraphics*[width=\textwidth]{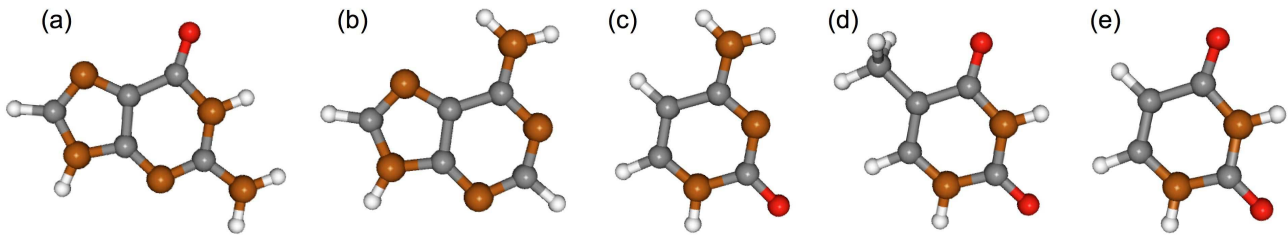}
\caption{(Color online) Schematic representation of the molecular structure 
of (a) guanine (G9K), (b) adenine, (c)  cytosine (C1), (d) thymine, and (e) uracil. 
Black, brown, red, white atoms are carbon, nitrogen, oxygen, and hydrogen, respectively.
The G9K and C1 notations for the guanine and cytosine tautomers are consistent 
with Ref.~\onlinecite{Bravaya10}.  }
\label{structures}
\end{center}
\end{figure*}

The determination of the ionization energies (IE),  electronic affinities (EA)
and character of the frontier orbitals of DNA and RNA nucleobases is
an important step towards a better understanding of the 
electronic properties and reactivity of nucleotides and nucleosides along the
DNA/RNA chains. Important phenomena such as nucleobases/protein interactions, 
defining the DNA functions \cite{polymerase}, or damages of the genetic material 
through oxidation or ionizing radiations \cite{Lodish}, are strongly related to 
these fundamental spectroscopic quantities.  Even though nucleobases in DNA/RNA 
strands are connected within the nucleotides to phosphate groups through a 
five-carbon sugar, several studies show that the highest-occupied orbital 
(the HOMO level) in nucleotides, which is responsible e.g. for the sensitivity
of the molecule to oxidation processes, remains localized on the nucleobases 
\cite{Close08}.  
Figure ~\ref{structures} shows the structures of the DNA and RNA nucleobases,
i.e.  the purines - adenine (A) and guanine (G),  and the pyrimidines - cytosine (C)
as well as thymine (T) in DNA and  uracil (U) in RNA.

Besides the overarching fundamental interest in understanding complex biological
processes at the microscopic level, \textit{ab initio} calculations of 
isolated nucleobases are interesting since recent high-level quantum
chemistry calculations \cite{Sanjuan06,Sanjuan08,Bravaya10}
allow to rationalize the rather large spread of experimental results 
concerning the electronic properties of the nucleobases in the gas phase 
\cite{Hush75,Dougherty78,Choi05,Trofimov06,Schwell08,Zaytseva09,Kostko10},
in particular as due to the existence of several isomers for
guanine and cytosine \cite{Bravaya10}.
Thus, these molecules offer a valuable mean to explore the merits of the 
so-called GW formalism \cite{Hedin65,Strinati80,Hybertsen86,Godby88,Onida02}
for isolated organic molecules, along the line of recent systematic  
studies of small  molecules \cite{Rostgaard10} or molecules such as fullerenes or 
porphyrins of interest for electronic or photovoltaic applications 
\cite{Dori06,Tiago08,Palumno09,Umari09,Stenuit10,Blase10}.

In the present work, we study by means of first-principles GW calculations
the quasiparticle properties of the DNA and RNA nucleobases, namely
guanine, adenine, cytosine, thymine and uracil.  We show in particular that
the GW correction to the Kohn-Sham eigenvalues brings the ionization
energies in much better agreement with experiment and high-level quantum
chemistry calculations. 
These results demonstrate the importance of self-consistency on the eigenvalues 
when performing GW calculations in molecular systems starting from (semi)local
DFT functionals, and the merits of a simple 
scheme based on a G$_0$W$_0$ calculation starting from Hartree-Fock like eigenvalues.

The GW approach is a Green's function formalism usually derived within
a functional derivative treatment \cite{Hedin65,Schwinger59}
allowing to prove that the 
two-body Green's function ($G_2$), involved in the equation of motion of 
the one-body time-ordered Green's function $G$, can be recast into a non-local
and energy-dependent self-energy operator 
$\Sigma(\mathbf{r},\mathbf{r'}|E)$. 
This self-energy $\Sigma$ accounts for exchange and correlation
in the present formalism. Since it is energy-dependent, it must be
evaluated at the $E={\varepsilon}^{QP}_i$ quasiparticle energies, 
where (i) indexes the molecular energy levels.
This self-energy involves 
$G(\mathbf{r},\mathbf{r'}|\omega)$, the dynamically-screened Coulomb potential 
$W(\mathbf{r},\mathbf{r'}|\omega)$,
and the so-called vertex correction $\Gamma$. A set of exact self-consistent 
(closed) equations connects $G$, $W$, $\Gamma$,  
and the independent-electron/full  polarisabilities
$\chi_0(\mathbf{r},\mathbf{r'}|\omega)$ and 
$\chi(\mathbf{r},\mathbf{r'}|\omega)$, respectively. In the GW
approximation (GWA), the three-body vertex operator $\Gamma$ is set to
unity, yielding the following expression for the self-energy:
\begin{eqnarray*}
 \Sigma(\mathbf{r},\mathbf{r'}|E) &=& { i \over 2\pi } \int d\omega \, e^{i\omega 0^+} 
      G(\mathbf{r},\mathbf{r'}| E+\omega) W(\mathbf{r},\mathbf{r'}|\omega) \\
\tilde{W}(\mathbf{r},\mathbf{r'}|\omega) &=& \int d\mathbf{r_1}
   d\mathbf{r_2} \, v(\mathbf{r},\mathbf{r_1}) \chi_0(\mathbf{r_1},\mathbf{r_2}|\omega)
     W(\mathbf{r_2},\mathbf{r'}|\omega), \\
 \chi_0(\mathbf{r},\mathbf{r'}|\omega) &=& \sum_{i,j} (f_i-f_j)
  {  \phi_i^*(\mathbf{r}) \phi_j(\mathbf{r}) \phi_j^*(\mathbf{r'}) \phi_i(\mathbf{r'})
     \over \varepsilon_i - \varepsilon_j + \omega \pm i\delta }
\end{eqnarray*}
where $v(\mathbf{r},\mathbf{r'})$ is the bare (unscreened) Coulomb potential
and ${\tilde  W} = W - v$. The $(\varepsilon_i,\phi_i)$ are ``zeroth-order" 
one-body eigenstates. Following the large bulk of work \cite{Onida02} devoted 
to GW calculations in solids, surfaces, graphene, nanotubes, or nanowires,
we use here Kohn-Sham DFT-LDA eigenstates.  It is shown below, and 
in Refs.~\onlinecite{Rostgaard10,Kaasbjerg10,Blase10,Hahn05}, that Hartree-Fock 
(or hybrid) solutions may constitute better starting points for molecular systems.  
$(f_i,f_j)$ are Fermi-Dirac occupation numbers, 
and $\delta$ an infinitesimal such that the poles of $W$ fall 
in the second and fourth quadrants of the complex plane.
In the GW approximation, the self-energy operator can be loosely interpreted
as a generalization of the Hartree-Fock method by replacing the bare Coulomb
potential with a dynamically screened Coulomb interaction accounting both for 
exchange and (dynamical) correlations.
An important feature of the GW approach is that not only ionization
energies and electronic affinities can be calculated, but also the full quasiparticle
spectrum. Further, both localized and infinite systems can be treated on the same
footing with long and short range screening automatically accounted for in the
construction of the screened Coulomb potential $W$.
More details about the present implementation can
be found in Ref.~\onlinecite{Blase10}.

Our calculations are based on a recently developed implementation of the GW
formalism (the {\sc Fiesta} code) using a gaussian auxiliary basis to expand the 
two-point operators such as the Coulomb potential, the susceptibilities or the 
self-energy \cite{Blase10}.  Dynamical correlations are included explicitly through 
contour deformation techniques. We start with a ground-state DFT calculation
using the {\sc Siesta} package \cite{siesta} and a large triple-zeta with double 
polarization (TZDP) basis \cite{KSbasis}. We fit the radial part of the numerical basis generated 
by the {\sc Siesta} code by up to five contracted gaussians in order to facilitate the 
calculation of the Coulomb matrix elements and of the matrix elements $\langle \phi_i|\beta|\phi_j \rangle$ 
of the auxiliary basis ($\beta$) between Kohn-Sham states. Such a scheme allows 
to exploit the analytic relations for the products of gaussian orbitals centered 
on different atoms or for their Fourier transform \cite{Blase10}.
Our auxiliary basis for first row elements is the tempered basis \cite{Cherkes09}
developed by Kaczmarski and coworkers \cite{Kaczmarski10}. Such a basis was tested
recently in a systematic study of several molecules of interest for photovoltaic 
applications \cite{Blase10}. Four gaussians for each l-channel with localization
coefficients $\alpha$=(0.2,0.5,1.25,3.2) a.u. are used for the (\textit{s,p,d}) channels
of C, O, and N atoms, while three gaussians with $\alpha$=(0.1,0.4,1.5) a.u.  describe
hydrogen \cite{largerbasis}.

\begin{figure}
\begin{center}
\includegraphics*[width=0.45\textwidth]{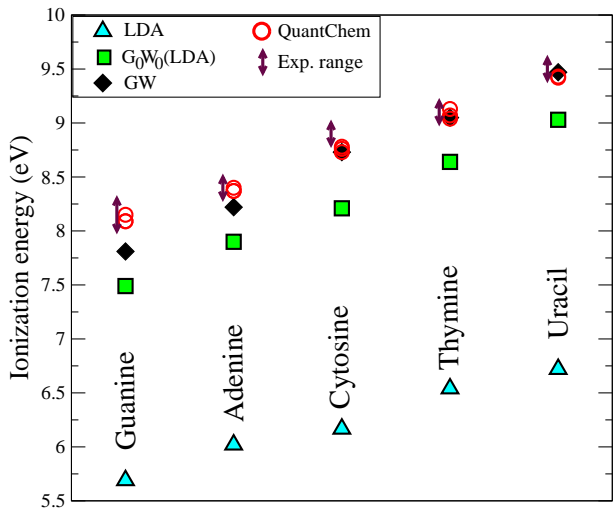}
\caption{(Color online) Ionization energies in eV. The vertical (maroon) error bars indicate the experimental
range. Triangles up (light blue): LDA values; (green) squares: G$_0$W$_0$(LDA) values; full black diamond: GW values;
(red) empty circles (QuantChem abbreviation): quantum chemistry, namely CCSD(T), CASPT2 and EOM-IP-CCSD, values 
(see text).  }
\label{iefig}
\end{center}
\end{figure}

\textbf{Ionization energies.}
We now comment on the values of the calculated first ionization energy (IE) as compiled in the
Table and Fig.~\ref{iefig}.
The comparison to the experimental data is complicated by the 0.2-0.3 eV range spanned by
the various experimental reports (vertical arrows Fig.~\ref{iefig}).
An additional complication in the case of cytosine and guanine, beyond the intrinsic difficulties 
in accurately measuring ionization energies in the gas phase, is that several gas phase tautomers 
exist \cite{Bravaya10} which differ from the so-called C1-cytosine
and G9K-guanine isomers commonly found in DNA (see Fig.~\ref{structures}).
State-of-the-art  \textit{ab initio} quantum chemistry calculations, namely coupled-cluster CCSD(T)
and multiconfigurational perturbation  (CASPT2) methods \cite{Sanjuan06,Sanjuan08},
studied the nucleobase tautomers that can be found along the DNA/RNA strands.
More recently, equation of motion coupled-cluster techniques (EOM-IP-CCSD) were performed on 
several isomers \cite{Bravaya10}.
All methods agree to within 0.04 eV for the average IE of the A, G, C, T 
tautomers we consider here, with a maximum discrepancy of 0.09 eV in the case of thymine. 
The  CASPT2 and CCSD(T) calculations agree to within 0.03 eV for all molecules.  These theoretical 
IE are commonly considered as the most reliable references and land within the experimental
error bars, except for the cytosine (C1) case where the calculated IEs are slightly smaller than 
the experimental lower bound \cite{IPisomers} (see Table and Fig.~\ref{iefig}).


Clearly, the ionization energy within DFT-LDA, as given by the negative 
HOMO Kohn-Sham level energy, significantly underestimates the IE by an average of 
$\sim$2.5 eV (29$\%$) \cite{PBE}.  The self-energy correction at the G$_0$W$_0$(LDA) 
level improves very significantly the situation by bringing the error 
to an average 0.5 eV (5.7$\%$) as compared to state-of-the-art quantum chemistry results. 
However, as emphasized in recent papers \cite{Rostgaard10,Kaasbjerg10,Blase10,Hahn05}, 
the overscreening induced by starting with LDA eigenvalues, which dramatically underestimate 
the band gap, tends to produce too small ionization energies. This problem can be solved at 
least partly by performing a simple self-consistency on the eigenvalues. We shall refer to this
approach as GW henceforth.
Such a self-consistency on the eigenvalues leads to a much reduced average error of 0.11 eV 
($\sim$1.3$\%$) as compared to the quantum chemistry reference. This good agreement 
certainly indicates the reliability of the present GW scheme for such systems. 
As shown in Fig.~\ref{iefig}, the largest discrepancies are observed for guanine and 
adenine (the purines), while the agreement is excellent for the three remaining bases.

In recent work, it was shown that for small molecules a non-self-consistent G$_0$W$_0$ 
calculation starting from Hartree-Fock eigenstates leads for the ionization energy to 
better results than a full self-consistent GW calculation where the wavefunctions 
are updated as well \cite{Rostgaard10,Kaasbjerg10}.
Consistent with this observation, a simple scheme 
relying on an Hartree-Fock-like approach was successfully tested on  silane, disilane,
and water \cite{Hahn05}, and larger molecules such as fullerenes or porphyrins 
\cite{Blase10}. In this ``G$_0$W$_0$ on Hartree-Fock (HF)" \textit{ansatz}, the input 
eigenvalues ($\tilde{\epsilon}_n$) are computed within a diagonal first-order perturbation theory where the 
DFT exchange-correlation contribution to the eigenvalues is replaced by the Fock 
exchange integral, namely:
\[ 
\tilde{\epsilon}_n = \epsilon^\mathrm{LDA}_n + < \psi^\mathrm{LDA}_n | \Sigma_x - V^\mathrm{LDA}_{xc} | \psi^\mathrm{LDA}_n>.
\]
where $\Sigma_x$ is the Fock operator. This approach, labeled G$_0$W$_0$(HF$_\mathrm{diag}$)
in the Table, produces an average error of 0.22 eV ($\sim$2.6$\%$). This good agreement with
both the GW and quantum chemistry calculations clearly speaks in favor of this simple scheme 
for molecular systems, or the full G$_0$W$_0$(HF) calculations tested in 
Ref.~\onlinecite{Rostgaard10}, which also avoids seeking self-consistency. 
A difficult issue lying ahead concerns e.g. hybrid systems, such as semiconducting surfaces
grafted by organic molecules, for which it is not quite clear what should be the best starting
point.

\begin{figure}[b]
\begin{center}
\includegraphics*[width=0.45\textwidth]{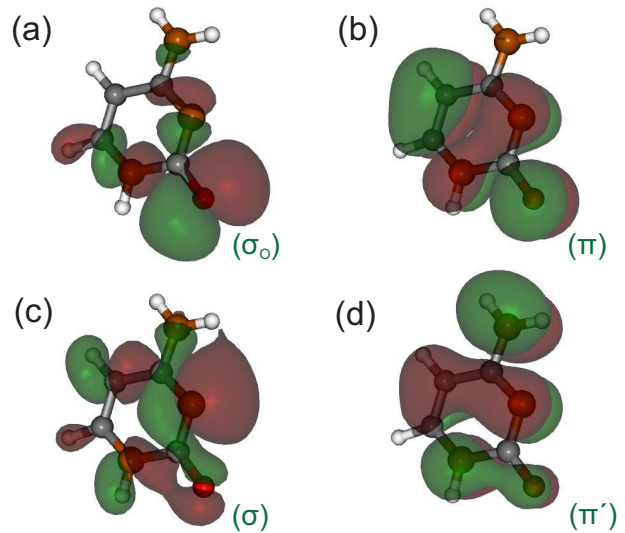}
\caption{(Color online)  Isodensity surface plot of the HOMO (${\sigma}_O$),  HOMO-1 
(${\pi}$), HOMO-2 (${\sigma}$), and HOMO-3 (${\pi}'$) LDA Kohn-Sham eigenstates of 
cytosine. Within GW, the ordering of states becomes ${\pi}$,${\pi}'$,${\sigma}_O$,
${\sigma}$ for HOMO to HOMO-3 (see text). }  
\label{waves}
\end{center}
\end{figure}

\begin{table*}
\begin{tabular}{||p{0.11\textwidth}|p{0.10\textwidth}|p{0.10\textwidth}|p{0.09\textwidth}|p{0.12\textwidth}|p{0.15\textwidth}|p{0.09\textwidth}|p{0.16\textwidth}||}
\hhline{|t:========:t|}
\multicolumn{8}{||c||}{\textbf{ Vertical ionization energies and vertical electronic affinities }} \\
\hhline{||--------||}
           &  LDA-KS &  G$_0$W$_0$(LDA) & GW   & G$_0$W$_0$(HF$_\mathrm{diag}$)  & CAS$^{a,b}$/CC$^{a,b}$ & EOM$^c$ & Experiment$^{d,e,f,g}$ \\
\hhline{||--------||}
G-LUMO    &  1.80     &  -1.04    &  -1.58    &  -1.77     &  -1.14$^a$/ &   &   \\
G-HOMO    &  5.69     &  7.49    &  7.81    &   7.76    & 8.09$^{b}$/8.09$^{b}$  & 8.15  &8.0-8.3$^d$   \\
G-HOMO-1   & 6.34     &  8.78    &  9.82    &   9.78    &  9.56$^{b}$/     & 9.86  & 9.90$^g$    \\
\hline
A-LUMO     &  2.22   &  -0.64   &  -1.14   &  -1.30     & -0.91$^a$/ &   & -0.56 to -0.45$^e$   \\
A-HOMO     &  6.02   &   7.90   &  8.22    &   8.23    & 8.37$^{b}$/8.40$^{b}$ &  8.37   &8.3-8.5$^d$, 8.47$^f$    \\
A-HOMO-1  &   6.28   &   8.75   &  9.47    &   9.51     & 9.05$^{b}$/    &  9.37   &  9.45$^f$ \\
\hline
C-LUMO  &  2.57              &  -0.45                &  -0.91               &  -1.05     & -0.69$^a$/-0.79$^a$ &    & -0.55 to -0.32$^e$   \\
C-HOMO     &  6.167 (${\sigma}_O$) &   8.21 ($\pi$)        & 8.73 ($\pi$)         & 9.05 ($\pi$)     &  8.73$^b$ ($\pi$)/8.76$^{b}$  &  8.78 ($\pi$) & 8.8-9.0$^d$, 8.89$^f$    \\
C-HOMO-1  &   6.172   ($\pi$)      &   8.80 (${\sigma}_O$) & 9.52 ($\pi$')        & 9.87 ($\pi$')    &  9.42$^b$ (${\sigma}_O$)/   &  9.54 ($\pi$')  & 9.45$^g$, 9.55$^f$       \\
C-HOMO-2  &   6.806 (${\sigma}$)    &  8.92  ($\pi$')       & 9.89 (${\sigma}_O$)  & 10.36 (${\sigma}_O$) &  9.49$^{b}$ ($\pi$')/   & 9.65 (${\sigma}_O$)  &         9.89$^f$    \\
C-HOMO-3  &  6.809  (${\pi}'$)    & 9.38 ($\sigma$)       & 10.22 (${\sigma}$)  & 10.64 (${\sigma}$) &  9.88$^{b}$(${\sigma}$)/    & 10.06 (${\sigma}$)  &         11.20$^f$    \\
\hline
T-LUMO    &   2.83   &  -0.14     &  -0.67     & -0.77      & -0.60$^a$/-0.65$^a$ &    &  -0.53 to -0.29$^e$   \\
T-HOMO    &   6.54   &   8.64     &   9.05     &   9.05    & 9.07$^{b}$/9.04$^{b}$ &    9.13       &  9.0-9.2$^d$, 9.19$^f$    \\
T-HOMO-1  &   6.68   &   9.34     &   10.41     &  10.40      &  9.81$^{b}$/    &  10.13      &  9.95-10.05$^d$,10.14$^f$   \\
\hline
U-LUMO    & 3.01                & -0.11               & -0.64                & -0.71                & -0.61$^a$/-0.64$^a$     &   &-0.30 to -0.22$^e$      \\
U-HOMO    & 6.72 (${\sigma}_O$) & 9.03 ($\pi$)        &  9.47 ($\pi$)        &  9.73 ($\pi$)        &  9.42$^{b}$ ($\pi$)/9.43$^b$  &  & 9.4-9.6$^d$    \\
U-HOMO-1  & 6.88  ($\pi$)       & 9.45 (${\sigma}_O$) & 10.54 (${\sigma}_O$) & 10.96 (${\sigma}_O$) &  9.83$^{b}$  (${\sigma}_O$)/  & & 10.02-10.13$^{d}$   \\
U-HOMO-2  & 7.55  (${\sigma}$)  &  9.88 ($\pi$')      & 10.66 ($\pi$')      & 11.06 ($\pi$')       &  10.41$^b$ ($\pi$')/     &   & 10.51-10.56$^d$  \\
U-HOMO-3  &  7.66 (${\pi}$')  & 10.33 ($\sigma$)      & 11.48 ($\sigma$)     &  11.90  ($\sigma$)  &  10.86$^b$ ($\sigma$)/   &   & 10.90-11.16$^{d}$  \\
\hhline{|b:========:b|}
 MAE LUMO &   3.29  &  0.33   &   0.18   &  0.31    &      &     &        \\
 MAE HOMO &   2.5  &  0.5   &  0.11    &   0.22   &          &           &        \\
\hhline{|b:========:b|}
\end{tabular}
\caption{Vertical ionization energies  and electronic affinities in eV as obtained from the negative
Kohn-Sham eigenvalues (LDA-KS), from non-self-consistent G$_0$W$_0$(LDA) calculations, 
from a GW calculation with self-consistency on the eigenvalues (GW), 
and from a non-self-consistent G$_0$W$_0$(HF$_\mathrm{diag}$)  calculation starting from
Hartree-Fock-like eigenvalues.  The $\sigma$ or $\pi$ character of the wavefunctions is indicated when the
GW correction changes the level ordering as compared to DFT-LDA (see text).
The acronyms CAS, CC and EOM stand for CASPT2,  CCSD(T) and equation of motion coupled-cluster
 high-level many body quantum chemistry calculations, respectively.
Theoretical values are reported for the C1-cytosine and G9K-guanine, while the experimental values average
over several tautomers.
The MAE is the mean absolute error in eV as compared to the quantum chemistry
reference calculations in columns 6 and 7.
$^a$Ref.~\onlinecite{Sanjuan08}.
$^b$Ref.~\onlinecite{Sanjuan06}.  
$^c$Ref.~\onlinecite{Bravaya10}.
$^d$Compiled in Ref.~\onlinecite{Sanjuan06}. 
$^e$Compiled in Ref.~\onlinecite{Sanjuan08}.
$^f$Ref.~\onlinecite{Trofimov06}.
$^g$Ref.~\onlinecite{Dougherty78}.
}
\label{tableIE}
\end{table*}

Next, we address the character of the HOMO level of cytosine and uracil.
It changes from DFT-LDA to GW calculations. We plot
in Fig.~\ref{waves}(a-d) the C1-cytosine DFT-LDA Kohn-Sham HOMO to (HOMO-3) eigenstates. The LDA HOMO level 
is an in plane $\sigma$ state with a strong component on the (\textit{p}$_x$,\textit{p}$_y$) oxygen
orbitals.  Such a state is labeled ${\sigma}_O$ in the Table and in the following.
The (HOMO-1) level is a more standard $\pi$-state with weight on the oxygen (\textit{p}$_z$)
orbital and a delocalized benzene ring $\pi$ molecular orbital.  Within the G$_0$W$_0$(LDA), GW and 
G$_0$W$_0$(HF$_\mathrm{diag}$) approaches, the LDA HOMO ${\sigma}_O$ state is pushed to a significantly 
lower energy and the $\pi$ state becomes the HOMO level. This level crossing brings the
GW calculations in agreement with many-body quantum chemistry calculations, which all predict
the $\pi$ state to be the HOMO level.
The same level crossing is observed in the case of uracil with the LDA HOMO and (HOMO-1) levels
being ${\sigma}_O$ and $\pi$-states respectively, while all GW results and quantum chemistry
calculations predict a reverse ordering.  
Our interpretation is that the very localized  ${\sigma}_O$ state suffers much more from the
spurious LDA self interaction than the rather delocalized  $\pi$ state. Even though it would be
wrong to reduce the dynamical GW self-energy operator to a self-interaction free functional,
the GW  correction certainly cures in part this well-known problem. 
The other bases, namely guanine, adenine, and thymine, all show
the correct $\pi$-character for the HOMO level.

The HOMO to (HOMO-1) energy difference averages to 0.80 eV and 1.12 eV within CASPT2 and 
EOM-IP-CCSD, respectively. Clearly, the average LDA energy spacing of 0.22 eV is significantly too 
small.  We find that the 0.77 eV G$_0$W$_0$(LDA) average value is close to the CASPT2 results, 
while the larger 1.29 eV GW result falls closer to the EOM-IP-CCSD energy difference. 
Averaging over all isomers, the experimental HOMO to (HOMO-1) energy spacing comes to
0.97 eV, in between the G$_0$W$_0$(LDA) or CASPT2 results and the GW or EOM-IP-CCSD values.
Even though it is too early for final conclusions about the merits of the various approaches, 
it seems fair to state that the LDA value is significantly too small, and that the situation 
is improved significantly by the GW correction.



\textbf{Electronic affinities.}
We conclude this study by exploring the electronic affinity (EA) of the nucleobases. 
They are provided in the Table as the negative sign of the LUMO Kohn-Sham energies.
Experimental data for guanine are missing. Further, the CASPT2 
and CCSD(T) results \cite{Sanjuan08} are clearly larger (in absolute value) 
than the highest experimental estimates. 
While again part of the discrepancy may come from the presence of several tautomers
in the gas phase, it certainly results as well from the fact that the electronic 
affinity is negative. A detailed discussion on the experimental difficulties in probing 
unbound states is presented in Ref.~\onlinecite{Bravaya10}. Taking again the CCSD(T) and 
CASPT2 calculations \cite{Sanjuan08} as a reference, the GW electronic affinities are quite 
satisfying, with a MAE of 0.18 eV. Such an agreement is rather impressive since the LDA 
electronic affinities show the wrong sign, with a discrepancy as compared to CASPT2 ranging 
from 2.9 eV to 3.6 eV. 
We observe that while the G$_0$W$_0$ EAs are smaller (in absolute value) than the
quantum chemistry ones, the GW EAs are larger. This contrasts with the IE case
where both G$_0$W$_0$ and GW values were smaller (see Fig. 2). Similar to the
quantum chemistry case, the GW values are found to systematically overestimate
the experimental results. Further study is needed to understand such a discrepancy
between theoretical and available experimental results.

In conclusion, we have studied on the basis of \textit{ab initio} GW calculations the ionization
energies and electronic affinities of the  DNA and RNA nucleobases, guanine, adenine, cytosine,
thymine and uracil. While a standard G$_0$W$_0$(LDA) calculation yields ionization energies that are
0.5 eV away from  CCSD(T)/CASPT2 reference quantum chemistry calculations, self-consistency on the 
eigenvalues brings the agreement to an excellent 0.11 eV average absolute error. A simple
G$_0$W$_0$ calculation starting from Hartree-Fock-like eigenvalues, avoiding the need for
self-consistency, shifts the agreement to 0.22 eV.  
The possibility of bringing the calculated values to within 0.1-0.2 eV from state-of-the-art
reference calculations with a scheme, the GW formalism, which allows to treat both finite size and
extended systems with a N$^4$ scaling, and permits to obtain the full quasiparticle spectrum,
paves the way to further studies of larger DNA strands and biological systems in general.
 
\textbf{Acknowledgements.}
C.F. is indebted to the European Union Erasmus program for funding.  Calculations 
have been performed on the CIMENT platform in Grenoble thanks to the Nanostar RTRA project.



\begin{thebibliography}{150bliography}
\bibitem{KSbasis}
It was shown in Refs.~\onlinecite{Rostgaard10,Blase10} for a large set of molecules
that a standard double-zeta  plus polarization (DZP) basis to expand the Kohn-Sham
eigenstates yields already IE within $\sim$ 0.1 eV as compared to the much larger 
TZDP basis. This certainly indicates that the present TZDP basis is an extremely 
good basis for such calculations.
\bibitem{largerbasis}
We have tested the use of a larger and more diffuse even-tempered basis with decay coefficients
$\alpha$=(0.15, 0.32, 0.69, 1.48, 3.2) a.u. for C, O, and N atoms. We find that the GW ionization
energies and electronic affinities change by 0.08 eV and 0.015 eV (MAE) respectively 
with no systematic trend.
\bibitem{IPisomers}
The EOM-IP-CCSD calculations predict that the IP of the most stable guanine 
(G7K) and cytosine (C2b) isomers (the most abundant in the gas phase) are 
0.14 eV and 0.08 eV larger than the IP of the G9K and C1 isomers we study.  Our 
calculations yields differences of 0.08 eV and 0.04 eV respectively (GW value). 
For cytosine, our GW calculations predict that the largest IE is that of the 
(C3a) tautomer which lands 0.19 eV above that of (C1).  This is consistent 
with the 0.2 eV experimental range, even though larger
than the 0.12 eV difference reported in Ref.~\onlinecite{Bravaya10} 
(EOM-IP-CCSD calculations).  Clearly, averaging over all isomers
would bring our results in better agreement with experiment. We note however that such
an average would require to know the abundance of each tautomer at the experimental
temperature. This stands beyond the purpose of the present paper.
\end{thebibliography}
\end{document}